\documentclass[graybox]{svmult}

\usepackage{type1cm}        % activate if the above 3 fonts are
\usepackage{makeidx}         % allows index generation
\usepackage{graphicx}        % standard LaTeX graphics tool
\usepackage{multicol}        % used for the two-column index
\usepackage[bottom]{footmisc}% places footnotes at page bottom

\usepackage{array}
\newcolumntype{C}[1]{>{\centering\let\newline\\\arraybackslash\hspace{0pt}}m{#1}}

\usepackage{newtxtext}       %
\usepackage{newtxmath}       % selects Times Roman as basic font

\usepackage{url}

\usepackage{tikz}
\usetikzlibrary{arrows.meta}

\makeindex             % used for the subject index
\begin{document}

\title*{From the Bloch sphere to phase space representations with the Gottesman-Kitaev-Preskill encoding}
\titlerunning{From the Bloch sphere to phase space representations with the GKP encoding} %for an abbreviated version of
\author{L. Garc{\'i}a-{\'A}lvarez, A. Ferraro and G. Ferrini}
\institute{L. Garc{\'i}a-{\'A}lvarez \at Department of Microtechnology and Nanoscience (MC2), Chalmers University of Technology, SE-412 96 G\"{o}teborg, Sweden, \email{lauraga@chalmers.se}
\and A. Ferraro \at Centre for Theoretical Atomic, Molecular and Optical Physics, Queen's University Belfast, Belfast BT7 1NN, United Kingdom
\and G. Ferrini \at Department of Microtechnology and Nanoscience (MC2), Chalmers University of Technology, SE-412 96 G\"{o}teborg, Sweden}
\maketitle

\abstract*{In this work, we study the Wigner phase-space representation of qubit states encoded in continuous variables (CV) by using the Gottesman-Kitaev-Preskill (GKP) mapping. We explore a possible connection between resources for universal quantum computation in discrete-variable (DV) systems, i.e. non-stabilizer states, and negativity of the Wigner function in CV architectures, which is a necessary requirement for quantum advantage. In particular, we show that the lowest Wigner logarithmic negativity of qubit states encoded in CV with the GKP mapping corresponds to encoded stabilizer states, while the maximum negativity is associated with the most non-stabilizer states, $H$-type and $T$-type quantum states.}

\abstract{In this work, we study the Wigner phase-space representation of qubit states encoded in continuous variables (CV) by using the Gottesman-Kitaev-Preskill (GKP) mapping. We explore a possible connection between resources for universal quantum computation in discrete-variable (DV) systems, i.e. non-stabilizer states, and negativity of the Wigner function in CV architectures, which is a necessary requirement for quantum advantage. In particular, we show that the lowest Wigner logarithmic negativity corresponds to encoded stabilizer states, while the maximum negativity is associated with the most non-stabilizer states, $H$-type and $T$-type quantum states.}


\section{Introduction}
\label{sec:intro}
Quantum computers, i.e. quantum devices in which information can be encoded, processed and read out, are predicted to solve certain computational problems faster than classical computers~\cite{shor1999}. Specifically, a problem is said to be hard to solve if its solution requires a number of steps  exponential in the size of the input, while polynomial time solutions are called efficient. An example of a problem believed to be hard to solve classically that can be efficiently solved by a quantum computer is factorization. While known classical algorithms factorize integer numbers in a time which scales exponentially with the size of the integer to factor, a quantum algorithm exists that only requires a polynomial time.

This technologically appealing property is referred to as \emph{quantum advantage}, and has recently motivated the undertaking of a global effort towards building a quantum computer. However, a conclusive experimental evidence of quantum advantage for computation is still lacking, since it has not yet been possible to build a quantum computer with enough elementary components to practically beat classical machines. Furthermore, the ultimate origin of quantum advantage is still unclear.

The traditional approach to encode information in quantum systems, based on two-level quantum systems with finite-dimensional Hilbert spaces, i.e. qubits, is an example of the discrete-variable (DV) approach. An alternative approach for information encoding uses continuous variables (CVs), i.e quantized variables with a continuous spectrum, such as the amplitude ($q$) and phase ($p$) quadratures of the quantized electromagnetic field, defined in an infinite-dimensional Hilbert space. Within this approach, one million optical modes have been entangled~\cite{yoshikawa2016,chen2014}. Beyond the optical realm, new CV implementations are studied in opto-mechanics~\cite{aspelmeyer2014} and with microwaves coupled to superconducting devices~\cite{ofek2016, wilson2011}, where high-order non-linearities can be engineered.

A fundamental tool for studying a classical dynamical system is the probability distribution on a phase space in which all possible states of the system are represented. Similarly, quantum systems can be conveniently and unambiguously described with quasi-probability distributions defined on the classical phase space~\cite{wigner1932,hillery1984,gibbons2004}. Although these useful mathematical constructs, such as the Wigner function, retain some properties of classical probability distributions, they can take negative values for quantum states.

A series of theorems has progressively narrowed down the characteristics that both DV and CV quantum computing architectures must possess in order to display quantum advantage.
In DV quantum information processors, the Gottesman-Knill theorem states that the so-called Clifford circuits, which are composed for example of Hadamard, $\pi/2$-phase and CNOT gates, when acting on stabilizer states, i.e. those generated with Clifford gates acting on the initial $n$-qubit register $|0\rangle_1\otimes|0\rangle_2\otimes \dots \otimes|0\rangle_n$, and followed by a Pauli measurement, can be efficiently simulated on a classical computer~\cite{gottesman1999,aaronson2004}. Pure states that are non-stabilizer are called magic, and are hence necessary to yield quantum advantage when acted on by Clifford circuits with Pauli measurements~\cite{bravyi2005}.
In CV quantum computation, it has been shown firstly that circuits with input, evolution and measurements solely described by Gaussian Wigner functions are efficiently simulatable by classical computers~\cite{bartlett2002}. Later it was shown that negativity of the Wigner function is a necessary requirement for quantum advantage, since quantum states and operations with positive Wigner functions (strictly including Gaussian circuits) can be classically efficiently simulated~\cite{mari2012}. Minimal extensions of positive Wigner function circuits that exhibit quantum advantage, where either the input, or the evolution, or the measurement are described by negative Wigner functions have been studied~\cite{chabaud2017,douce2017,hamilton2017,chakhmakhchyan2017,douce2019}. Finally, the criteria for efficient classical simulatability have been extended by using other phase-space representations, namely Husimi and Glauber-Sudarshan~\cite{rahimi-keshari2016}.

A bridge between the DV and the CV worlds is provided by CV-codes, i.e. by sets of CV states that allow for encoding DV states such that orthogonal wavefunctions represent different DV states. One such example is the Gottesman-Kitaev-Preskill (GKP) code, where the qubit logical states are encoded in trains of delta functions at different locations~\cite{gottesman2001}. The encoding of discrete quantum information into infinite-dimensional quantum systems is used to get a high-quality qubit protected from environmental noise~\cite{menicucci2014}. The GKP code is particularly suitable for our analysis since Clifford gates on the qubit encoded states are given by Gaussian operations, which in principle lead us to an analogy between DV and CV requirements for classical efficient simulatability of quantum operations.

In this manuscript, we analyze the negativity of the Wigner function for any single-qubit state mapped in CV architectures with the GKP code, with the aim of establishing a relation between DV and CV criteria for quantum advantage. In section~\ref{sec:GKP}, we review in detail the GKP code that we use in our work. In section~\ref{sec:wigner}, we compute the Wigner function of any single-qubit GKP encoded state, and we compare the results for encoded stabilizer and non-stabilizer states. In section~\ref{sec:WLN}, we quantify the negativity of the Wigner function for both cases, and we observe that stabilizer encoded states saturate the lower bound of negativity, while the most non-stabilizer states, also known as magic states, show the maximum amount of negativity.
We conclude in section~\ref{sec:conclusions} with our final remarks.

\section{GKP encoding of qubit states}
\label{sec:GKP}
The formal GKP encoding maps a qubit into an oscillator using non-normalizable superpositions of infinitely squeezed states in the position $q$ and momentum $p$ quadratures of the oscillator~\cite{gottesman2001}. We review the GKP qubit states used in this work, which are defined as
\begin{align}
 \label{eq:perfectGKP}
  & |0\rangle = \sum_{s=-\infty}^{\infty} |q=2\sqrt{\pi}s \rangle \nonumber \\
  & |1\rangle = \sum_{s=-\infty}^{\infty} |q=\sqrt{\pi}(1 +2s) \rangle ,
\end{align}
for which the wavefunction $\Psi(q) = \langle q|\Psi \rangle$ is a sum of delta functions, since $\langle q|q=x\rangle = \delta(x)$.

In practice, the qubit states must be normalizable, and thus are defined approximating the previous expression with finitely squeezed states, and weighting the infinite sum of squeezed states by a Gaussian envelope. The approximated states are quasi-orthogonal states given by
\begin{align}
 \label{eq:imperfectGKP}
  & |\bar{0}\rangle \propto \sum_{s=-\infty}^{\infty} \int_{-\infty}^{\infty} e^{-2 \pi \kappa^2 s^2} e^{-\frac{(q-2\sqrt{\pi}s)^2}{2\sigma^2}}|q\rangle dq \nonumber \\
  & |\bar{1}\rangle \propto \sum_{s=-\infty}^{\infty} \int_{-\infty}^{\infty} e^{-2 \pi \kappa^2 s^2} e^{-\frac{(q-(2s+1)\sqrt{\pi})^2}{2\sigma^2}}|q\rangle dq ,
\end{align}
with $\kappa^{-1}$, the width of the Gaussian envelope, and $\sigma$, the width of the Gaussian peaks substituting the delta functions. These imperfect GKP states are suitable for numerical computations but introduce a probability of error in the identification of $|\bar{0}\rangle$ and $|\bar{1}\rangle$. In our calculations, we use the perfect GKP states given in Eq.~(\ref{eq:perfectGKP}) for obtaining analytical results, and imperfect GKP states in Eq.~(\ref{eq:imperfectGKP}) for numerical results.

\section{Phase space Wigner representation of GKP encoded states}
\label{sec:wigner}
The Wigner function of a pure state $|\Psi\rangle$ is defined as
\begin{equation}
 \label{eq:wigner}
 W(q,p) \equiv \frac{1}{2\pi} \int_{-\infty}^{\infty} dx e^{ipx} \Psi\left(q+\tfrac{x}{2}\right)^{*}\Psi\left(q-\tfrac{x}{2}\right),
\end{equation}
with $\Psi(x) = \langle x|\Psi \rangle$ the wavefunction of the quantum system.

We consider infinitely squeezed GKP states, that is, the ideal logical qubit GKP states $|j\rangle$ with $j=0,1$ given in Eq.~(\ref{eq:perfectGKP}). The corresponding Wigner function reads~\cite{gottesman2001}
\begin{equation}
 \label{eq:wigner_diag}
 W_{j}(q,p) = \frac{1}{4\sqrt{\pi}} \sum_{st} (-1)^{st} \delta\left(p-\tfrac{\sqrt{\pi}}{2}s\right)\delta\left(q-\sqrt{\pi} j -\sqrt{\pi} t\right).
\end{equation}

\begin{figure*}[!hbt] \sidecaption
 \centering
 \begin{tikzpicture}[line cap=round, line join=round, >=Triangle]
  \clip(-2.19,-2.49) rectangle (4.58,2.58);
  \draw [shift={(0,0)}, lightgray, fill, fill opacity=0.0] (0,0) -- (56.7:0.4) arc (56.7:90.:0.4) -- cycle;
  \draw [shift={(0,0)}, lightgray, fill, fill opacity=0.0] (0,0) -- (-135.7:0.4) arc (-135.7:-33.2:0.4) -- cycle;
  \draw(0,0) circle (2cm);
  \draw [rotate around={0.:(0.,0.)},dash pattern=on 3pt off 3pt] (0,0) ellipse (2cm and 0.9cm);
  \draw (0,0)-- (0.70,1.07);
  \draw [->] (0,0) -- (0,2);
  \draw [->] (0,0) -- (-0.81,-0.79);
  \draw [->] (0,0) -- (2,0);
  \draw [dotted] (0.7,1)-- (0.7,-0.46);
  \draw [dotted] (0,0)-- (0.7,-0.46);
  \draw (-0.08,-0.43) node[anchor=north west] {$\phi$};
  \draw (0.01,0.9) node[anchor=north west] {$\theta$};
  \draw (-1.04,-0.72) node[anchor=north west] {$|+\rangle$};
  \draw (2.07,0.3) node[anchor=north west] {$|i\rangle$};
  \draw (-0.3,2.6) node[anchor=north west] {$|0\rangle$};
  \draw (-0.3,-2) node[anchor=north west] {$|1\rangle$};
  \draw (0.1,1.65) node[anchor=north west] {$|\Psi\rangle$};
  \scriptsize
  \draw [fill] (0,0) circle (1.5pt);
  \draw [fill] (0.7,1.1) circle (0.5pt);
 \end{tikzpicture}
 \caption{Geometrical representation of pure qubit states in the Bloch sphere.}
 \label{fig:bloch}
\end{figure*}
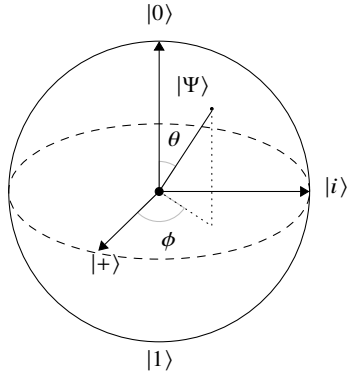

We now take into account arbitrary pure qubit states given by superpositions of GKP states as $|\Psi\rangle = \cos \tfrac{\theta}{2} |0\rangle + e^{i\phi} \sin \tfrac{\theta}{2} |1\rangle$, which can be represented in the surface of the Bloch sphere as shown in Fig.~\ref{fig:bloch}. The Wigner function for a qubit state depends consequently on the the angles $\theta, \phi$ of its Bloch sphere representation. It reads
\begin{align}
 \label{eq:wigner_genGKP}
 W(\theta,\phi;q,p) = & \frac{1}{2\pi} \int_{-\infty}^{\infty} dx e^{ipx} \bigg[ \cos^2 \tfrac{\theta}{2} \Psi_{0}\left(q+\tfrac{x}{2}\right)^{*}\Psi_{0}\left(q-\tfrac{x}{2}\right) \nonumber \\
                      & + \sin^2 \tfrac{\theta}{2} \Psi_{1}\left(q+\tfrac{x}{2}\right)^{*}\Psi_{1}\left(q-\tfrac{x}{2}\right) \nonumber                                                        \\
                      & + \cos \tfrac{\theta}{2} \sin \tfrac{\theta}{2} e^{i\phi} \Psi_{0}\left(q+\tfrac{x}{2}\right)^{*}\Psi_{1}\left(q-\tfrac{x}{2}\right) \nonumber                         \\
                      & + \cos \tfrac{\theta}{2} \sin \tfrac{\theta}{2} e^{-i\phi} \Psi_{1}\left(q+\tfrac{x}{2}\right)^{*}\Psi_{0}\left(q-\tfrac{x}{2}\right)\bigg] ,
\end{align}
with $\Psi_{i}$, $i=0,1$, the wavefunctions corresponding to the GKP states $|i\rangle$, $i=0,1$. A detailed derivation can be found in appendix~1. Explicitly, we have
%in appendix~\ref{app:wigner}
\begin{align}
 \label{eq:wignerGKP}
 W(\theta,\phi;q,p) = & \cos^2 \tfrac{\theta}{2} W_{0}(q,p) + \sin^2 \tfrac{\theta}{2} W_{1}(q,p) \nonumber                                                                                                          \\
                      & + \frac{\sin \theta}{4\sqrt{\pi}} \sum_{st} (-1)^{st} \cos\left(\phi +s\tfrac{\pi}{2}\right) \delta\left(q-\tfrac{\sqrt{\pi}}{2}(1+2t)\right)  \delta\left(p-\tfrac{s\sqrt{\pi}}{2}\right) ,
\end{align}
which can be pictured in a grid of square cells of $\Delta q=\Delta p=\tfrac{\sqrt{\pi}}{2}$. By analyzing Eq.~(\ref{eq:wigner_diag}) and Eq.~(\ref{eq:wignerGKP}), we thus observe that the Wigner function consists of a sum of delta functions positioned at all the sites of the lattice in phase space with coordinates $(l,m)\equiv (q=l\tfrac{\sqrt{\pi}}{2},p=m\tfrac{\sqrt{\pi}}{2})$ for $l$ and $m$ integer numbers. The coefficients for each site are given by
\begin{align}
 \label{eq:wignerGKPcoef}
 w_{l m}(\theta,\phi) = \left\{
 \begin{array}{ll}
  \frac{1}{4\sqrt{\pi}} \left(\cos^2 \tfrac{\theta}{2} + \sin^2 \tfrac{\theta}{2}\right) & \text{for } l \text{ even, } m \text{ even}    \\
  \frac{1}{4\sqrt{\pi}} \left(\cos^2 \tfrac{\theta}{2} - \sin^2 \tfrac{\theta}{2}\right) & \text{for } l=4u ,\, m \text{ odd}             \\
  \frac{1}{4\sqrt{\pi}} \left(\sin^2 \tfrac{\theta}{2} - \cos^2 \tfrac{\theta}{2}\right) & \text{for } l=4u+2 ,\, m \text{ odd}           \\
  \frac{1}{4\sqrt{\pi}} \sin \theta \cos \phi                                            & \text{for } \left\{ \begin{array}{l}
   l=4u+3 ,\, m=4v \\
   l=4u+1 ,\, m=4v \\
  \end{array}
  \right.                                                                                                                                 \\
  \frac{-1}{4\sqrt{\pi}} \sin \theta \cos \phi                                           & \text{for } \left\{ \begin{array}{l}
   l=4u+3 ,\, m=4v+2 \\
   l=4u+1 ,\, m=4v+2 \\
  \end{array}
  \right.                                                                                                                                 \\
  \frac{-1}{4\sqrt{\pi}} \sin \theta \sin \phi                                           & \text{for } \left\{ \begin{array}{l}
   l=4u+3 ,\, m=4v+3 \\
   l=4u+1 ,\, m=4v+1 \\
  \end{array}
  \right.                                                                                                                                 \\
  \frac{1}{4\sqrt{\pi}} \sin \theta \sin \phi                                            & \text{for } \left\{ \begin{array}{l}
   l=4u+3 ,\, m=4v+1 \\
   l=4u+1 ,\, m=4v+3 \\
  \end{array}
  \right.                                                                                                                                 \\
 \end{array}
 \right.
\end{align}
with $u$ and $v$ integer numbers.

\begin{figure*}[p!]
 \begin{tabular*}{\textwidth}{c @{\extracolsep{\fill}} c}
  \includegraphics[width=0.417\textwidth]{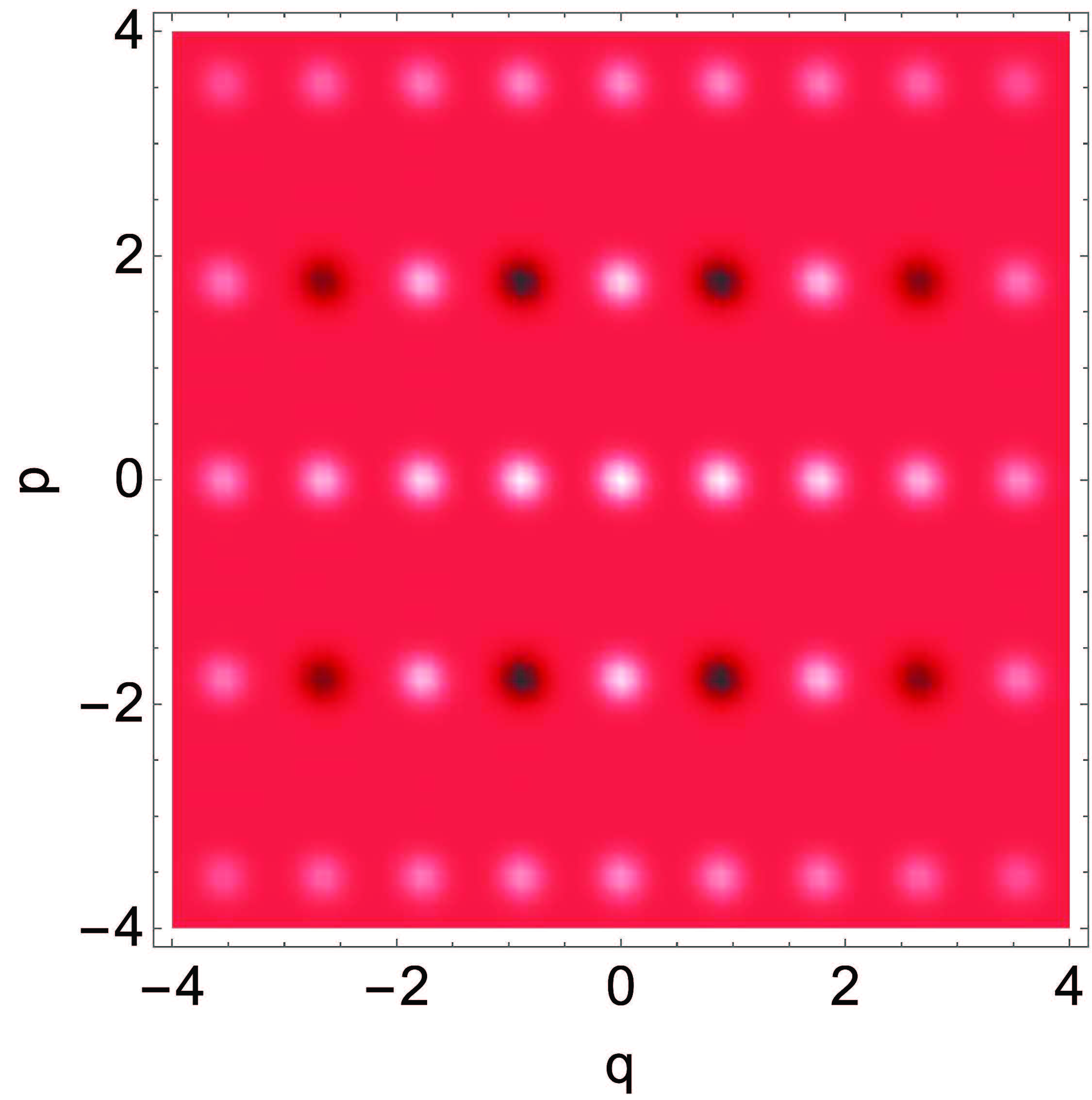} &
  \includegraphics[width=0.417\textwidth]{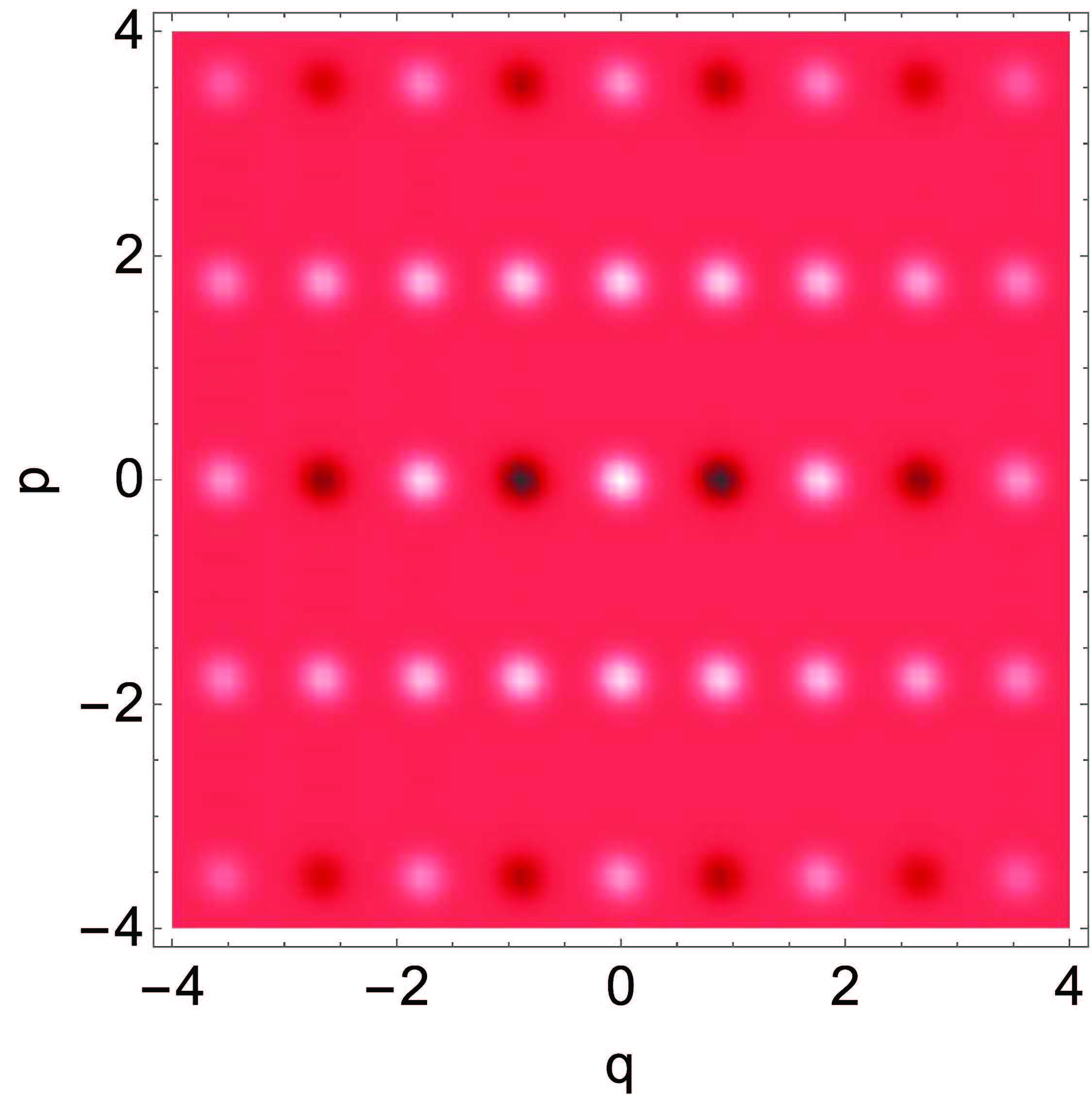} \\
  {\footnotesize (a) $|+\rangle= \frac{1}{\sqrt{2}}(|0\rangle +|1\rangle )$} &
  {\footnotesize (b) $|-\rangle= \frac{1}{\sqrt{2}}(|0\rangle -|1\rangle )$} \\
  {\footnotesize} & {\footnotesize} \\
  \includegraphics[width=0.417\textwidth]{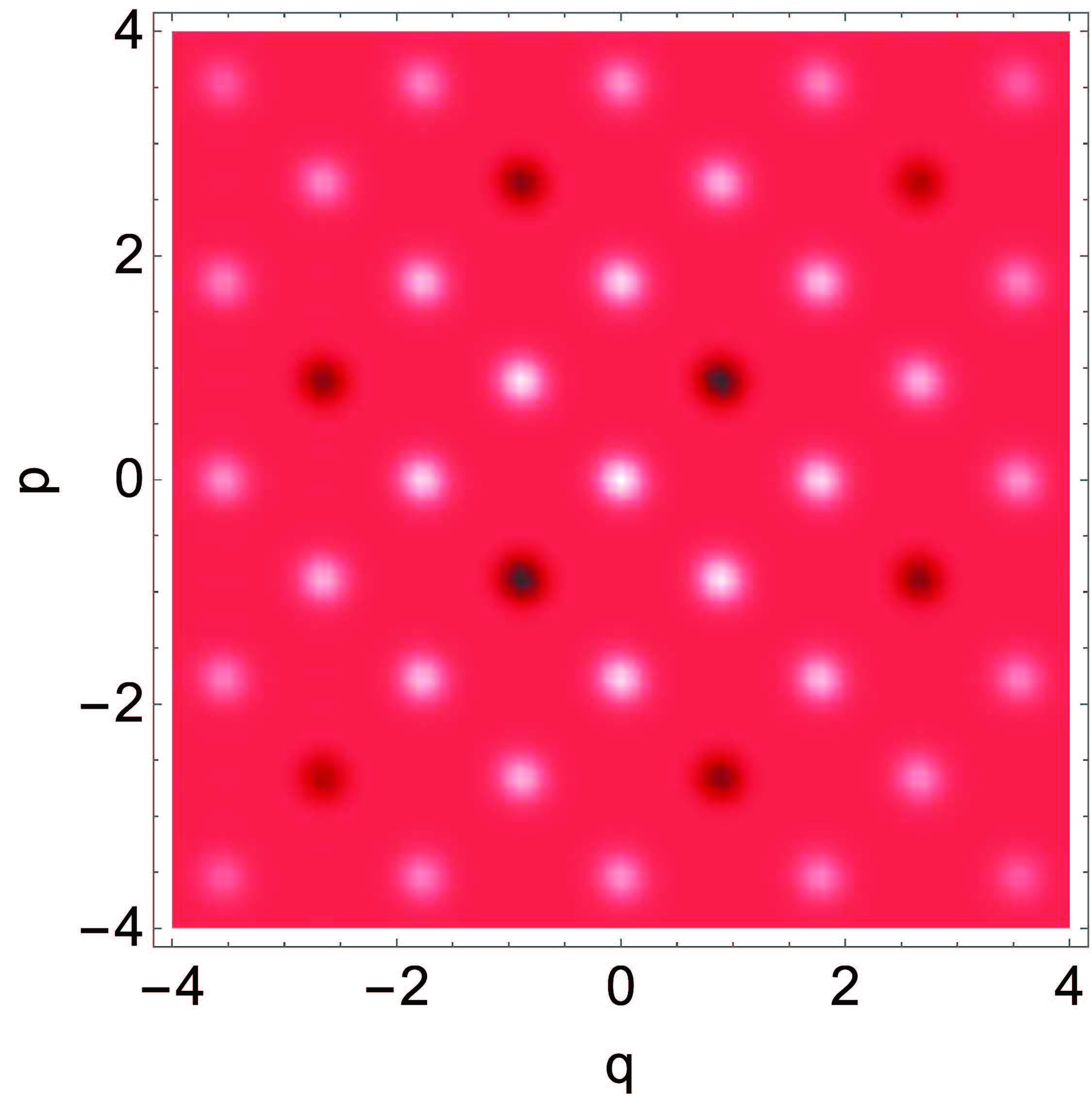} &
  \includegraphics[width=0.417\textwidth]{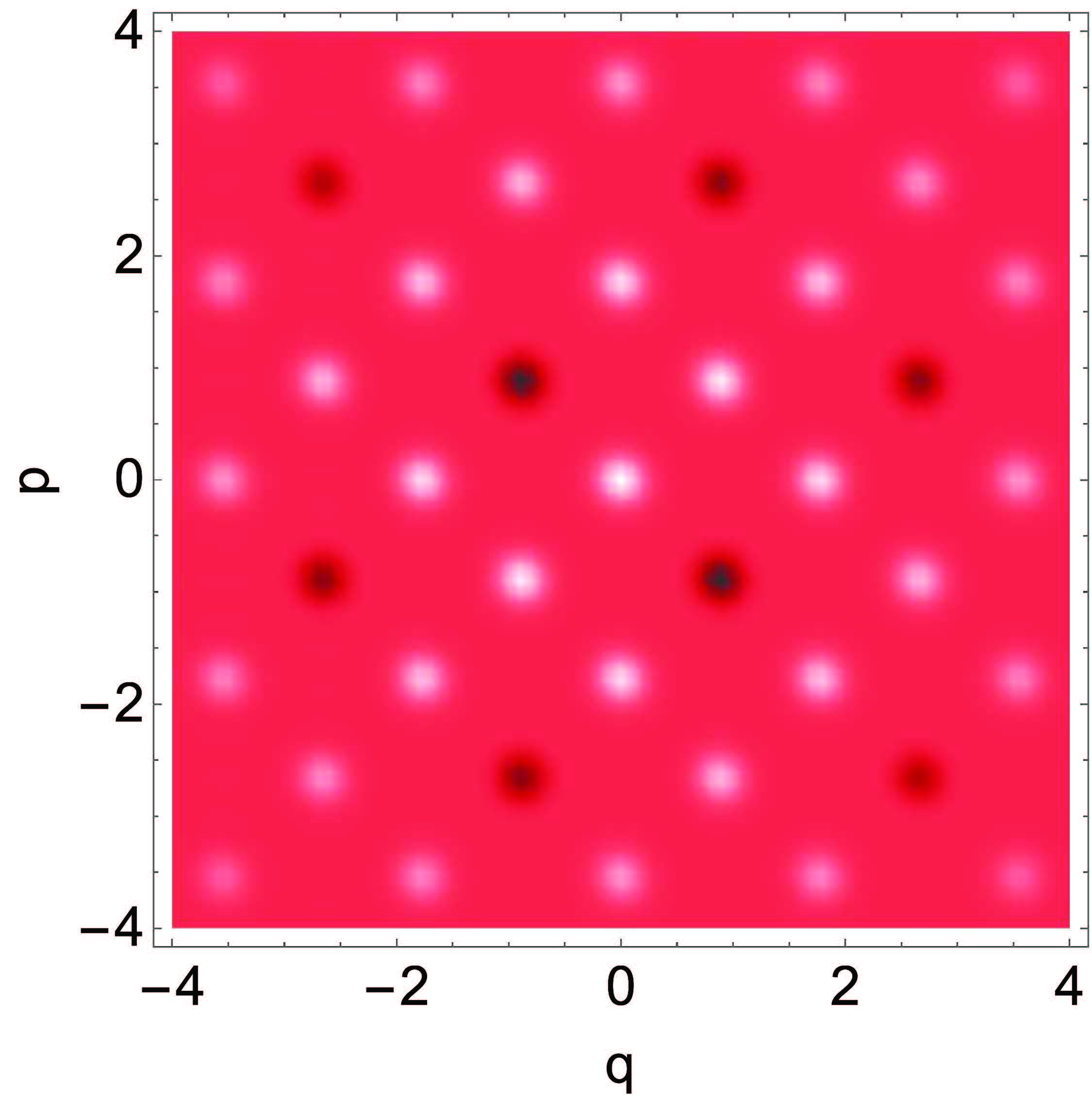} \\
  {\footnotesize (c) $|+i\rangle= \frac{1}{\sqrt{2}}(|0\rangle +i|1\rangle )$} &
  {\footnotesize (d) $|-i\rangle= \frac{1}{\sqrt{2}}(|0\rangle -i|1\rangle )$} \\
  {\footnotesize} & {\footnotesize} \\
  \includegraphics[width=0.417\textwidth]{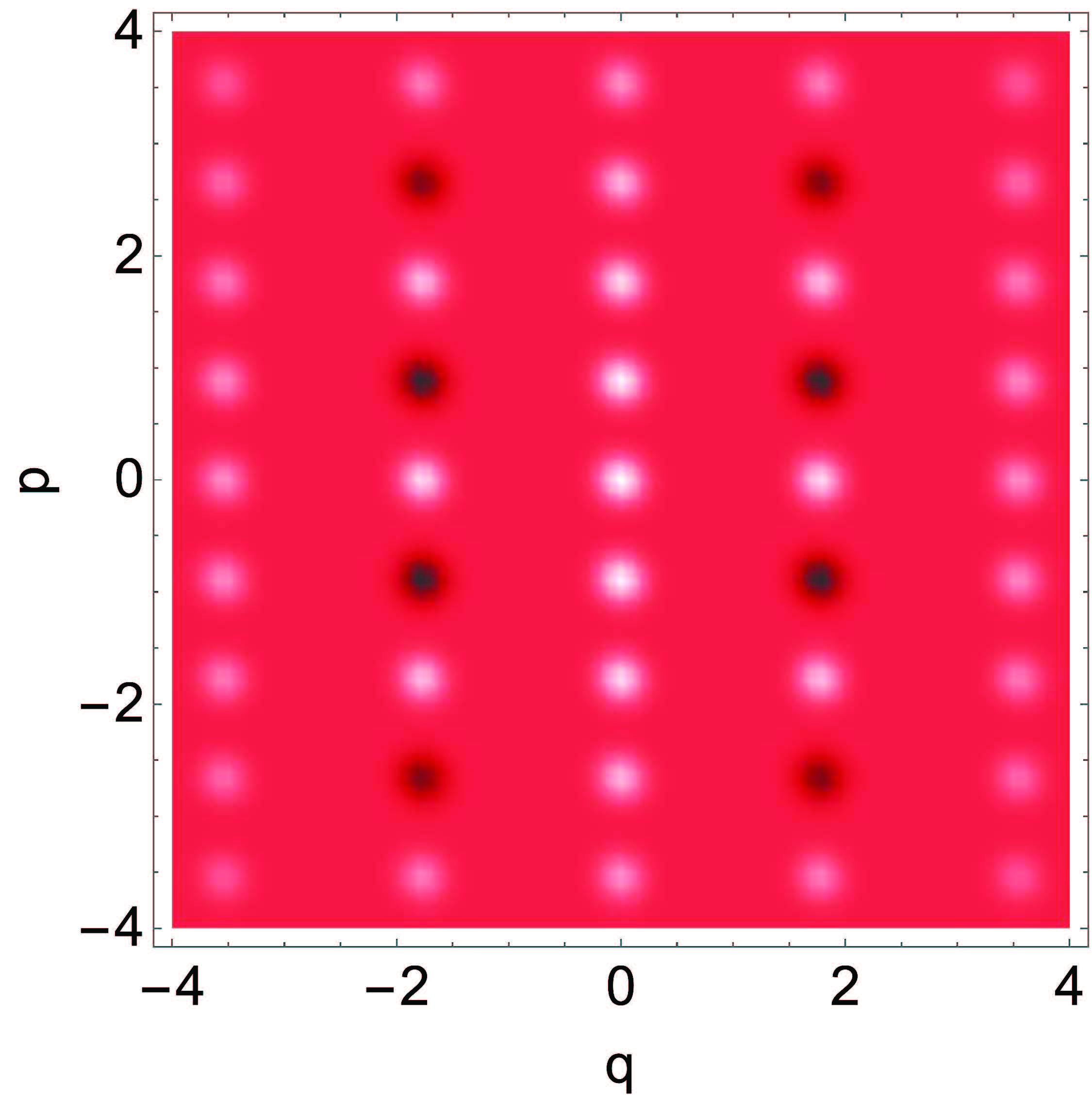} &
  \includegraphics[width=0.417\textwidth]{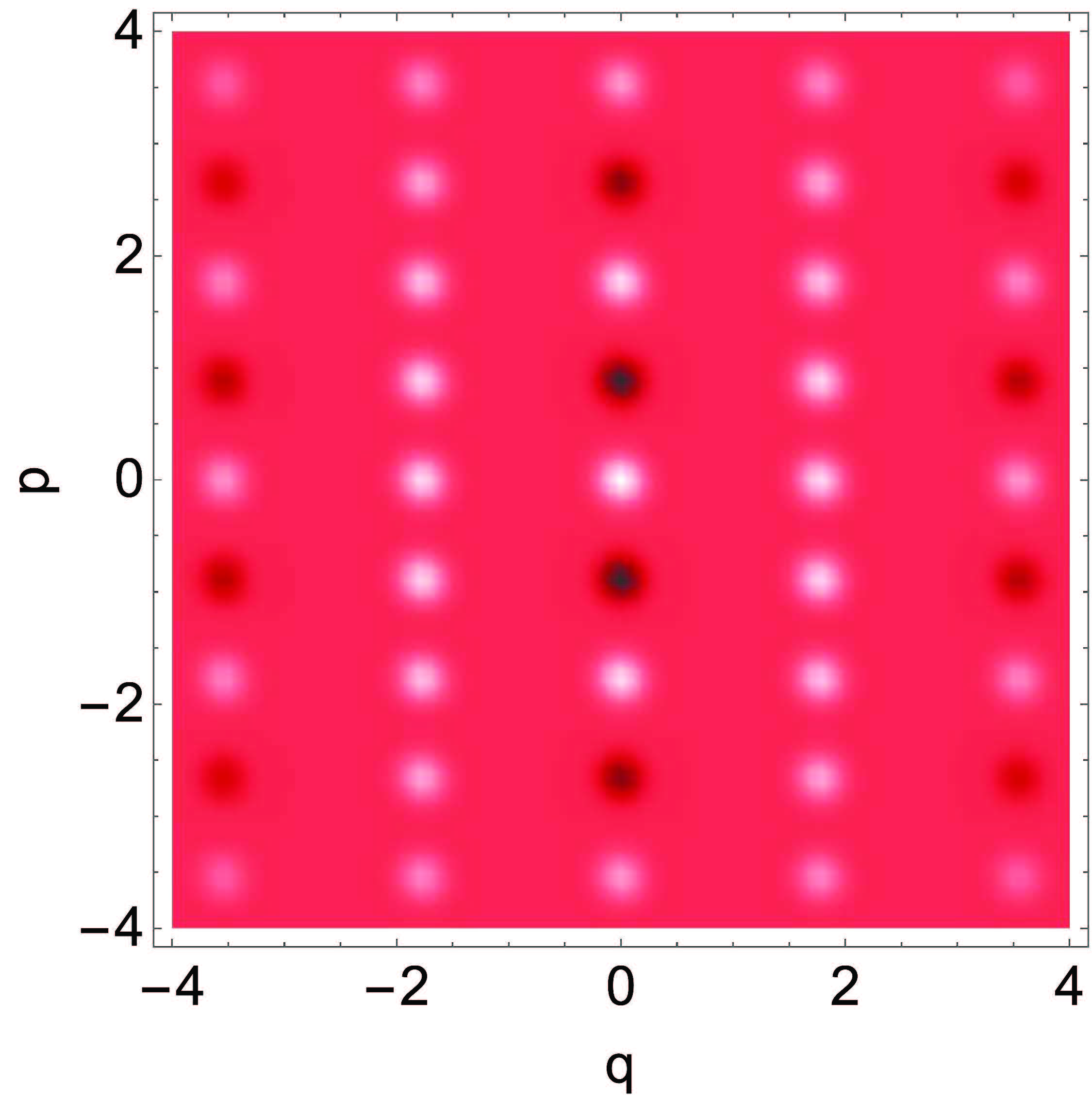} \\
  {\footnotesize (e) $|0\rangle$} &
  {\footnotesize (f) $|1\rangle$} \\
 \end{tabular*}
 \caption{Wigner function of qubit GKP encoded stabilizer states. The function acquires non-zero values on the dark and white peaks, where it has a negative value (dark) and positive value (white), respectively. We consider finitely squeezed states as in Eq.~(\ref{eq:imperfectGKP}), with $\sigma = \kappa = 0.2$.}
 \label{fig:wstab}
\end{figure*}

In particular, we consider the six single-qubit stabilizer pure states, corresponding to the eigenvectors of the Pauli matrices $\sigma_x$, $\sigma_y$, and $\sigma_z$,
\begin{align}
 \label{eq:stabilizer}
  & \sigma_x : \qquad |+\rangle= \frac{1}{\sqrt{2}}(|0\rangle +|1\rangle ) \qquad |-\rangle= \frac{1}{\sqrt{2}}(|0\rangle - |1\rangle ) \nonumber     \\
  & \sigma_y : \qquad |i\rangle= \frac{1}{\sqrt{2}}(|0\rangle + i|1\rangle ) \qquad |-i\rangle= \frac{1}{\sqrt{2}}(|0\rangle - i|1\rangle ) \nonumber \\
  & \sigma_z : \qquad |0\rangle \phantom{=\frac{1}{\sqrt{2}}(|+\rangle+|+\rangle)} \qquad |1\rangle .
\end{align}

The Wigner functions of single-qubit stabilizer states mapped in CV via the GKP code are shown in Fig.~\ref{fig:wstab}. We observe a similar pattern repeated periodically and isotropically in the whole phase space, with one quarter of negative delta functions with respect to the total amount of peaks. It is possible to obtain from the initial state $|0\rangle$ all stabilizer states with Clifford operations, which for a single qubit are generated in DV by the Hadamard $H$, and $\frac{\pi}{2}$-phase gates $R_{\frac{\pi}{2}}$,
\begin{align}
 \label{eq:CliffordDV}
  & H_{\phantom{\frac{\pi}{2}}} : \qquad |0\rangle \rightarrow |+\rangle , \qquad |1\rangle \rightarrow |-\rangle , \nonumber \\
  & R_{\frac{\pi}{2}} : \qquad |0\rangle \rightarrow |0\rangle , \qquad |1\rangle \rightarrow e^{i\frac{\pi}{2}}|1\rangle .
\end{align}

With the GKP encoding, these gates in CV correspond to the Fourier transform $F$, and the $\pi/2$-phase gate $P$, which are the symplectic transformations
\begin{align}
 \label{eq:CliffordCV}
  & F : \qquad q \rightarrow p , \qquad p \rightarrow -q , \nonumber \\
  & P : \qquad q \rightarrow q , \qquad p \rightarrow p - q .
\end{align}

Let us consider now the single-qubit magic states $|
 T\rangle$ and $|H\rangle$,
\begin{align}
 \label{eq:magic}
  & |T\rangle=\cos \tfrac{\theta}{2} |0\rangle+\sin \tfrac{\theta}{2} e^{i\frac{\pi}{4}} |1\rangle \quad \text{with} \quad \theta = \arccos\left(\tfrac{1}{\sqrt{3}}\right)  \nonumber \\
  & |H\rangle=\frac{1}{\sqrt{2}} \left(|0\rangle+ e^{i\frac{\pi}{4}} |1\rangle\right),
\end{align}
which are the maximal non-stabilizer states in the Bloch sphere and in the equatorial plane of the Bloch sphere, respectively~\cite{bravyi2005}. There are 8 $T$-type magic states and 12 $H$-type magic states, which can be obtained from the states in Eq.~(\ref{eq:magic}) with Clifford transformations (see Fig.~\ref{fig:abswneg}).

\begin{figure*}[hbt!]
 \begin{tabular*}{\textwidth}{c @{\extracolsep{\fill}} c}
  \includegraphics[width=0.417\textwidth]{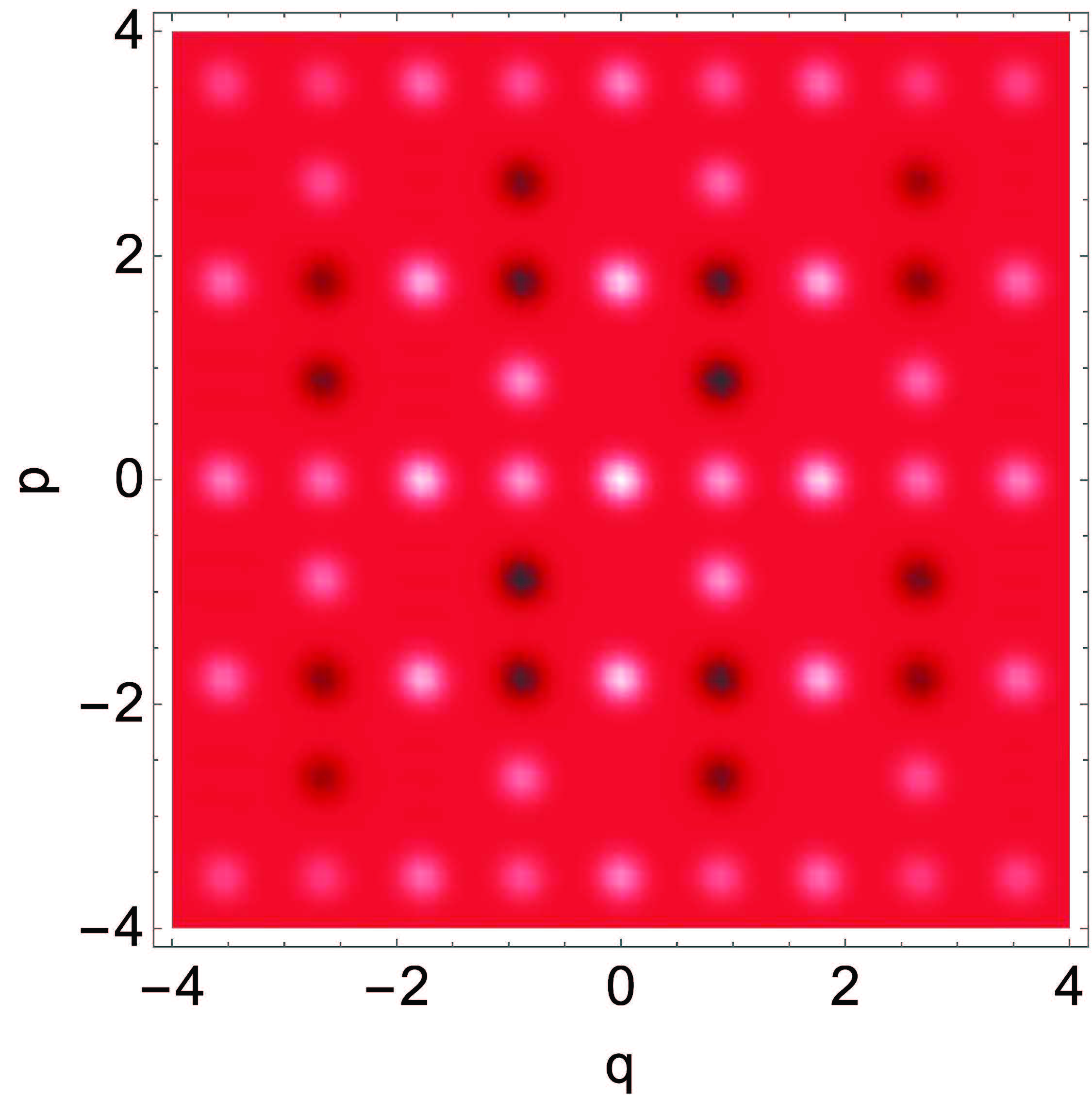} &
  \includegraphics[width=0.417\textwidth]{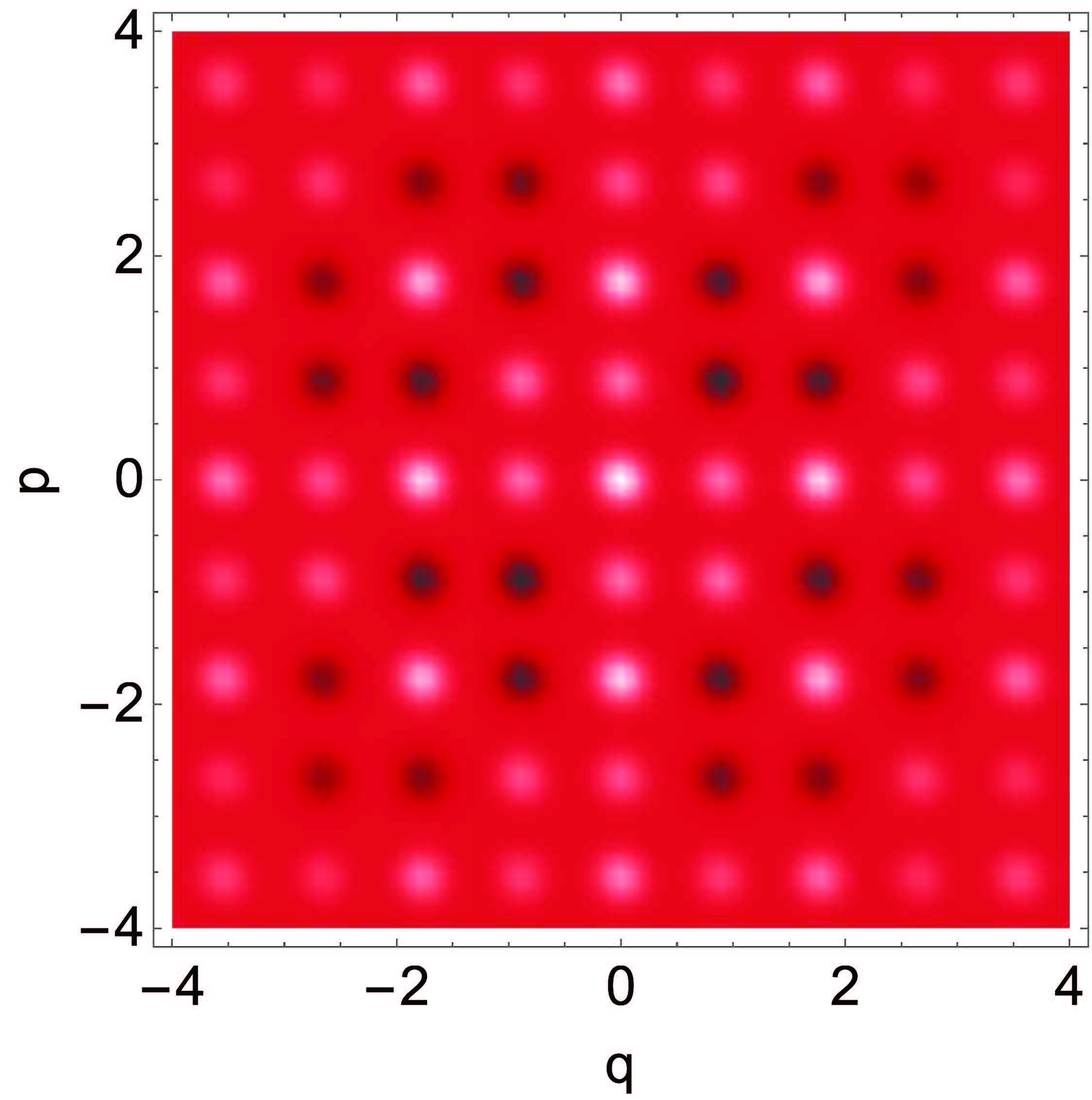} \\
  {\footnotesize (a) $|H\rangle$} &
  {\footnotesize (b) $|T\rangle$}
 \end{tabular*}
 \caption{Wigner function of qubit GKP encoded magic states. The function acquires non-zero values on the dark and white peaks, where it has a negative value (dark) and positive value (white), respectively. We consider finitely squeezed states as in Eq.~(\ref{eq:imperfectGKP}), with $\sigma = \kappa = 0.2$. (a)~$|H\rangle$ state, and (b)~$|T\rangle$ state, both given in Eq.~(\ref{eq:magic}).}
 \label{fig:wmagic}
\end{figure*}

The Wigner function of the quantum states $|T\rangle$ and $|H\rangle$ mapped in CV via the GKP code are shown in Fig.~\ref{fig:wmagic}. Both the numerical computations and the analytical expression indicate that the number of negative peaks increases with respect to the Wigner function of stabilizer states, although the proportion remains as before: one quarter of negative delta functions and three quarters of positives ones. As one can observe comparing  Fig.~\ref{fig:wstab} and Fig.~\ref{fig:wmagic}, it is not possible to obtain a non-stabilizer Wigner function pattern from a stabilizer one with single-qubit Clifford GKP encoded operations as those given in Eq.~(\ref{eq:CliffordCV}).

\section{Quantification of negativity of the Wigner function for GKP encoded states}
\label{sec:WLN}
We now aim at quantifying the volume of the negative part of the Wigner function for the different types of states that we have introduced. The quantification of the volume of the negative part of the Wigner function in CV is related to the monotone \emph{Wigner logarithmic negativity} (WLN)~\cite{kenfack2004,albarelli2018}, defined as
\begin{equation}
 \label{eq:WLN}
 {\mathcal{W}}(\rho) = \log_{2} \left( \int dq dp |W(q,p)|\right),
\end{equation}
with $W(q,p)$ the Wigner function of the state or operator $\rho$. The WLN has allowed for the derivation of a bound in the number of necessary copies of an input state for the conversion to a target state~\cite{albarelli2018}.

As we have already mentioned, the proportion of negative delta functions compared to positive ones in the Wigner function of both stabilizer and magic encoded states is one quarter. However, we observe in Fig.~\ref{fig:wstab} and Fig.~\ref{fig:wmagic} that the Wigner function of non-stabilizer states is composed of more peaks in the phase-space, resulting in a higher number of negative delta peaks. We now use the WLN for analysing the differences in both kind of states, since it tracks the amount of negativity instead of the proportion.

We consider the Wigner function of perfect GKP states in Eq.~(\ref{eq:wignerGKP}). The negativity takes an infinite value since the Wigner function has support in the whole phase space $\mathbb{R}^2$, but the delta functions are periodically arranged following symmetric patterns that are repeated along the two axes in a similar way for each qubit superposition state. Therefore, we may consider the same square unit cell of dimension $(\Delta q, \Delta p)=(2\sqrt{\pi},2\sqrt{\pi})$ for all cases, and compare the negativity within the same finite area in phase space. We choose the unit cell corresponding to $s=t=0$ in Eq.~(\ref{eq:wignerGKPcoef}), which contains sixteen delta functions given by $l$ and $m$ with values in the set $\{0,1,2,3\}$.

Explicitly, the Wigner function in the unit cell domain $q \in [0,2\sqrt{\pi} )$ and $p \in [0,2\sqrt{\pi} )$ is given by
\begin{equation}
 \label{eq:wignerGKPcell}
 W_{\rm{cell}} (\theta,\phi;q,p) = \sum_{l,m=0}^{3} w_{lm}(\theta,\phi) \delta\left(q-l\tfrac{\sqrt{\pi}}{2}\right)\delta\left(p-m\tfrac{\sqrt{\pi}}{2}\right),
\end{equation}
where the coefficients correspond to those defined in Eq.~(\ref{eq:wignerGKPcoef}).
The absolute value of the Wigner function for the unit cell can be taken as the absolute value of the summands, since for any coordinate $(q_i,p_i)$ in the domain only one of the terms is different from zero due to the properties of the delta functions. Thus,
\begin{equation}
 \label{eq:wignerGKPcellabs}
 |W_{\rm{cell}} (\theta,\phi;q,p)| = \sum_{l,m=0}^{3} |w_{lm}(\theta,\phi)| \delta\left(q-l\tfrac{\sqrt{\pi}}{2}\right)\delta\left(p-m\tfrac{\sqrt{\pi}}{2}\right) .
\end{equation}
As a result, the WLN corresponding to a unit cell in the phase space for any pure qubit GKP encoded state $|\Psi\rangle = \cos \tfrac{\theta}{2} |0\rangle + e^{i\phi} \sin \tfrac{\theta}{2} |1\rangle$ characterized in the Bloch sphere by angles $(\theta,\phi)$ is given by
\begin{align}
 \label{eq:WLNcelleq}
 {\mathcal{W}}_{\rm cell}(\theta,\phi) & = \log_{2} \left( \int dq dp |W_{\rm{cell}} (\theta,\phi;q,p)|\right) \nonumber                                                                                             \\
                                       & = \log_{2} \sum_{l,m=0}^{3} |w_{lm}(\theta,\phi)| \left( \int dq dp \delta\left(q-\tfrac{l\sqrt{\pi}}{2}\right)\delta\left(p-\tfrac{m\sqrt{\pi}}{2}\right)\right) \nonumber \\
                                       & = \log_{2} \sum_{l,m=0}^{3} |w_{lm}(\theta,\phi)| .
\end{align}
Explicitly, the WLN per cell of a qubit state is then given by
\begin{equation}
 \label{eq:WLNcell}
 {\mathcal{W}}_{\rm cell}(\theta,\phi) = \log_{2} \bigg[ \frac{1}{\sqrt{\pi}} \big[1 +\left|\cos^2 \tfrac{\theta}{2} - \sin^2 \tfrac{\theta}{2}\right| +\left|\sin \theta \cos \phi\right| +\left|\sin \theta \sin \phi\right| \big] \bigg].
\end{equation}

Now, we compare the finite WLN per cell, ${\mathcal{W}}_{\rm cell}$, for different magic and stabilizer state by analysing for simplicity the integral over a unit cell of the absolute value of the Wigner function $\int dq dp |W_{\rm{cell}}|$, i.e the argument of the logarithm in Eq.~(\ref{eq:WLNcelleq}). The corresponding values are provided in Table~\ref{tab:WLN}. We observe that the WLN per cell for GKP encoded qubit stabilizer states is lower than for non-stabilizer states.
 {\setlength{\tabcolsep}{1em}
  \begin{table}[!ht]
   \caption{Integral over a unit cell of the absolute value of the Wigner function for stabilizer states and magic states.}
   \centering
   \bgroup
   \def\arraystretch{1.5}
   \begin{tabular}{c|c|c|c|}
    \cline{2-4}
                                      & $\theta$                         & $\phi$  & $ \sqrt{\pi}\int|W_{\rm{cell}}|$ \\ \hline
    \multicolumn{1}{|l|}{$|0\rangle$} & $0$                              & $0$     & $2$                              \\ \hline
    \multicolumn{1}{|l|}{$|+\rangle$} & $\pi/2$                          & $0$     & $2$                              \\ \hline
    \multicolumn{1}{|l|}{$|i\rangle$} & $\pi/2$                          & $\pi/2$ & $2$                              \\ \hline
    \multicolumn{1}{|l|}{$|H\rangle$} & $\pi/2$                          & $\pi/4$ & $1 +\sqrt{2} \approx 2.41$       \\ \hline
    \multicolumn{1}{|l|}{$|T\rangle$} & $\arccos\left(1/\sqrt{3}\right)$ & $\pi/4$ & $1 +\sqrt{3} \approx 2.73$       \\ \hline
   \end{tabular}
   \egroup
   \label{tab:WLN}
  \end{table}}
Since all GKP encoded qubit states have a proportion of one quarter of negative delta functions, the WLN is different from zero for all of them. This Wigner negativity is intrinsic to the use of the GKP encoding, that is, it is only attributed to the fact that we are using an encoding where even the stabilizer states are represented by non-Gaussian wavefunctions exhibiting Wigner negativity. This intrinsic Wigner negativity in GKP states might be sufficient to promote Gaussian quantum circuits to universal quantum computation~\cite{baragiola2019}.

We now compute the lower bound of this intrinsic negativity by considering
\begin{equation}
 \int dq dp|W_{\rm{cell}} (\theta, \phi;q,p)| \geq \left|\int dq dp W_{\rm{cell}} (\theta, \phi;q,p)\right| = \frac{2}{\sqrt{\pi}} .
\end{equation}
We observe that stabilizer states saturate the lower bound of the integral over a unit cell of the absolute value of the Wigner function, $\int|W_{\rm{cell}}|$, and therefore they are the least negative qubit GKP encoded states.

\begin{figure*}[hbt!]
 \begin{tabular*}{\textwidth}{C{0.45\textwidth} C{0.45\textwidth}}
  \includegraphics[width=0.46\textwidth]{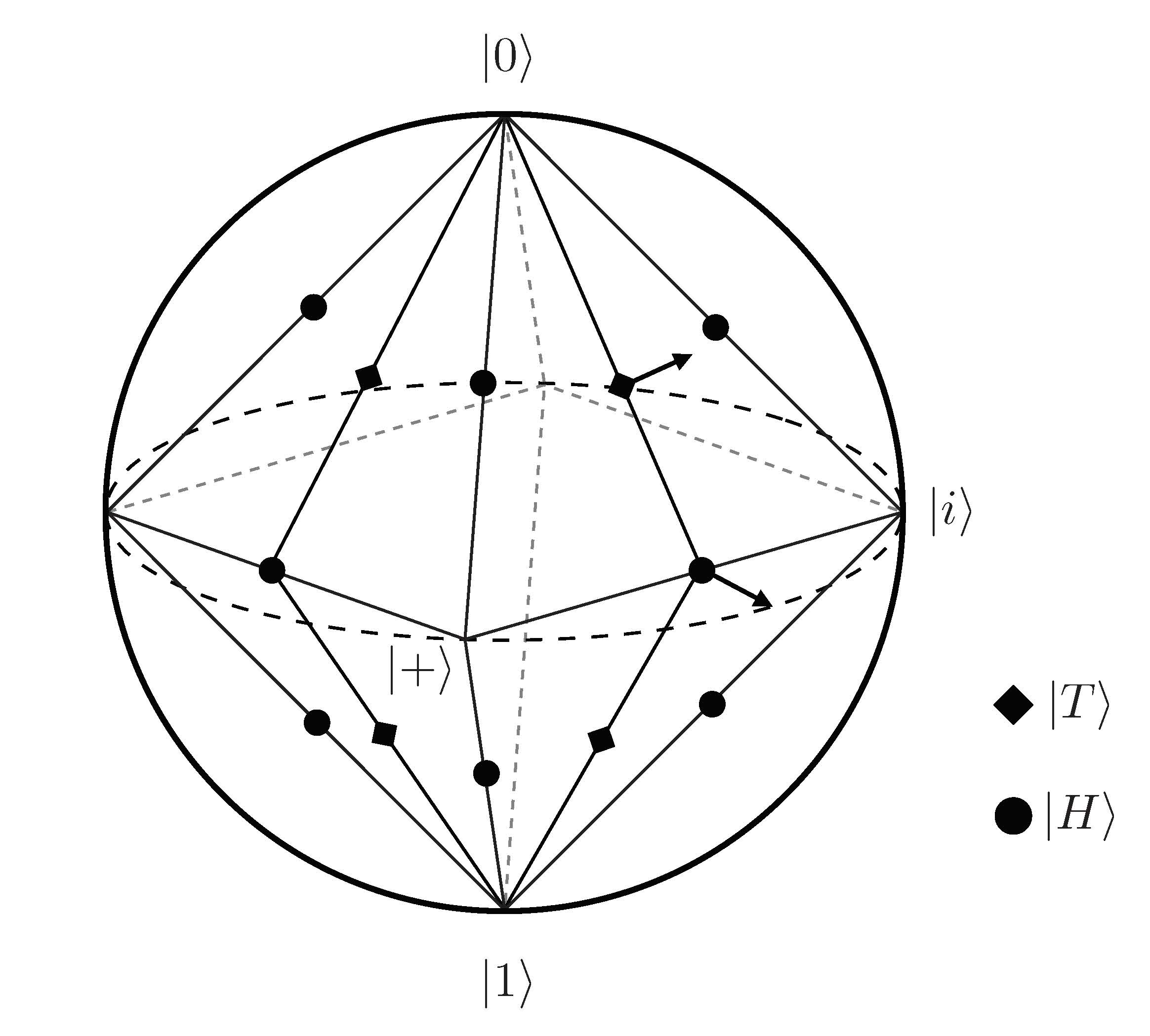} &
  \includegraphics[width=0.42\textwidth]{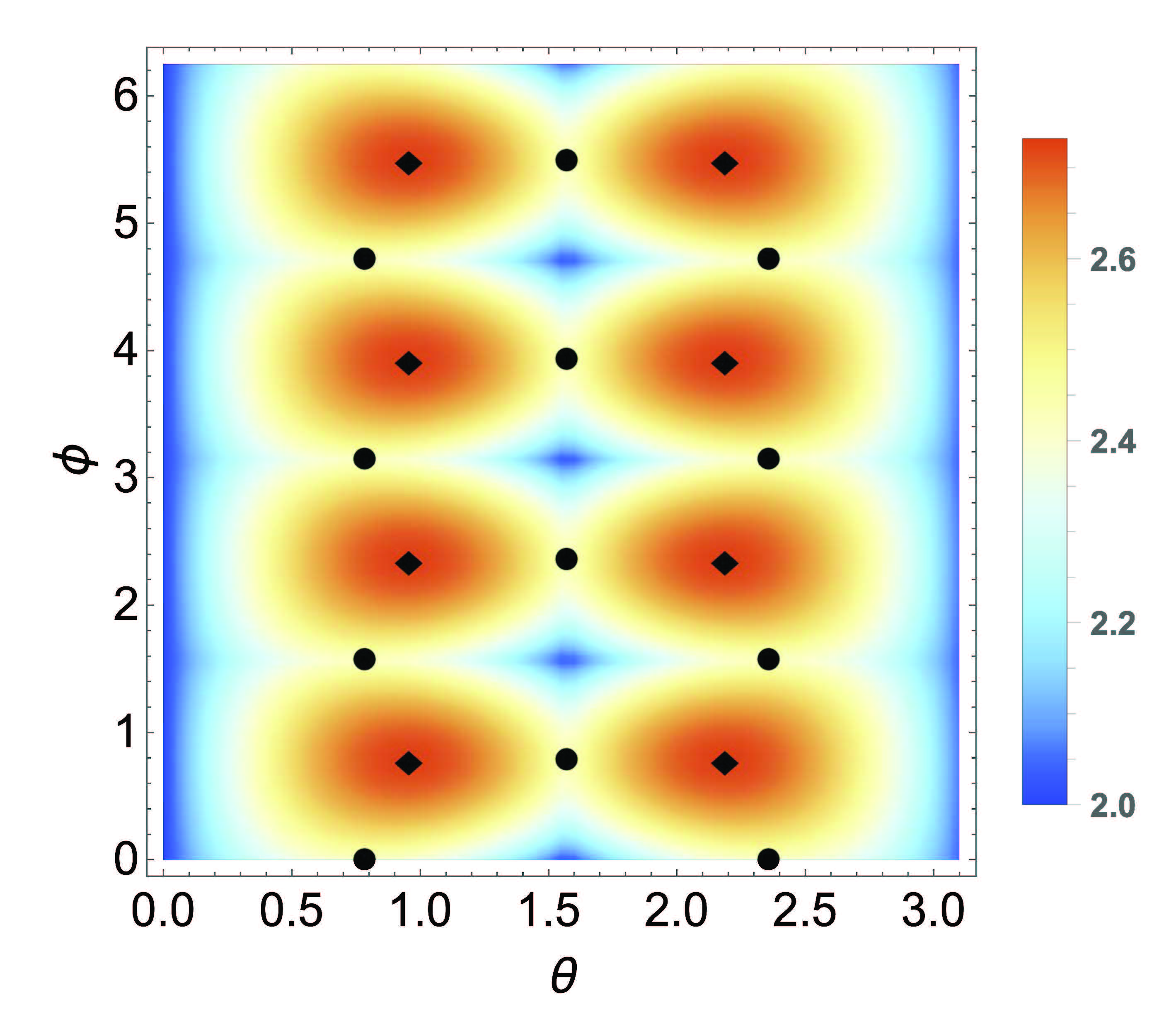} \\
  {\footnotesize (a) $T$-type and $H$-type magic states.} &
  {\footnotesize (b) $\sqrt{\pi}\int|W_{\rm{cell}}(\theta,\phi;q,p)|dq dp$ for single-qubit states.}
 \end{tabular*}
 \caption{(a) Representation of single-qubit states on the Bloch sphere. Stabilizer states correspond to the vertices of an octahedron embedded in the sphere. The most non-stabilizer states are those projected on the surface of the sphere from the middle points of the edges of the octahedron, $H$-type magic states (circle), and perpendicularly from the center of the faces, $T$-type magic states (diamond), as indicated by the arrows~\cite{bravyi2005}. (b) Quantification of negativity of the Wigner function of qubit GKP encoded states with $\sqrt{\pi}\int|W_{\rm{cell}}|$. We consider all qubit states, described by the angles $(\theta,\phi)$, with  $\theta \in [0,\pi)$ and $\phi \in [0,2\pi)$.}
 \label{fig:abswneg}
\end{figure*}

We show in Fig.~\ref{fig:abswneg} the function $\sqrt{\pi}\int|W_{\rm{cell}}(\theta,\phi;q,p)|dq dp$, which is proportional to the argument of the logarithm in the WLN. It is computed for all qubit states, characterized in the Bloch sphere with $(\theta,\phi)$, with $\theta \in [0,\pi)$ and $\phi \in [0,2\pi)$. We observe that the stabilizer states are the least negative, whereas the maxima appears for $|T\rangle$ qubit states, which are the most non-stabilizer single-qubit states. On the equatorial plane of the Bloch sphere (see Fig.~\ref{fig:bloch}), $\theta=\frac{\pi}{2}$, the maxima appears for $|H\rangle$ states, which are the most non-stabilizer states on that plane.

\section{Conclusions}
\label{sec:conclusions}

In this work, we use CV tools as the Wigner phase-space representation for studying DV single-qubit states encoded in infinite Hilbert spaces with the GKP mapping. We give an analytical expression for the Wigner function of any GKP encoded qubit state, and quantify the amount of negativity with the WLN.
All qubit states have non-zero WLN, and therefore we cannot distinguish which states and processes are classically efficiently simulatable with current criteria for quantum advantage in CV systems. On the other hand, our quantitave analysis of the WLN for GKP encoded states shows differences for stabilizer and non-stabilizer states, since the first ones are the least negative, saturating the lower bound of negativity. The most non-stabilizer states, $H$-type and $T$-type quantum states, reach the maximum negativity. Our results suggest a possible connection between a DV characterization of resources for universal quantum computation and CV necessary criteria for quantum advantage.

A natural perspective stemming from this work is to explore the relation between different states with non-zero WLN and the computational complexity of quantum circuits including these states.

Note added: Work presented in the International Symposium on Mathematics, Quantum Theory, and Cryptography (MQC), held in September 2019 in Fukuoka, Japan (https://www.mqc2019.org/mqc2019/program). After the acceptance of the present work in the Springer's ``Mathematics for Industry'' series, we became aware of an independent work in which the negativity of the Wigner function of GKP states is also analytically evaluated~\cite{yamasaki2019}.

\begin{acknowledgement}
 We thank P. Milman and A. Ketterer for sharing with us a Mathematica code that was useful in the explorative stage of this project. L. G.-\'{A}. and G. F. acknowledge support from the Wallenberg Center for Quantum Technology (WACQT), and G. F. acknowledges financial support from the Swedish Research Council through the VR project QUACVA.
\end{acknowledgement}

\section*{Appendix 1}
%\addcontentsline{toc}{section}{Appendix 1}
\label{app:wigner}
A detailed derivation of Eq.~(\ref{eq:wignerGKP}) is provided here. Firstly, we can conveniently rewrite the Wigner function in Eq.~(\ref{eq:wigner_genGKP}) as follows
\begin{align}
 \label{eq:appwigner}
 W(\theta,\phi;q,p) = & \frac{1}{2\pi} \int_{-\infty}^{\infty} dx e^{ipx} \bigg[ \cos^2 \tfrac{\theta}{2} \Psi_{0}\left(q+\tfrac{x}{2}\right)^{*}\Psi_{0}\left(q-\tfrac{x}{2}\right) \nonumber               \\
                      & + \sin^2 \tfrac{\theta}{2} \Psi_{1}\left(q+\tfrac{x}{2}\right)^{*}\Psi_{1}\left(q-\tfrac{x}{2}\right) \nonumber                                                                      \\
                      & + \cos \tfrac{\theta}{2} \sin \tfrac{\theta}{2} e^{i\phi} \Psi_{0}\left(q+\tfrac{x}{2}\right)^{*}\Psi_{1}\left(q-\tfrac{x}{2}\right) \nonumber                                       \\
                      & + \cos \tfrac{\theta}{2} \sin \tfrac{\theta}{2} e^{-i\phi} \Psi_{1}\left(q+\tfrac{x}{2}\right)^{*}\Psi_{0}\left(q-\tfrac{x}{2}\right)\bigg] \nonumber                                \\
 =                    & \cos^2 \tfrac{\theta}{2} W_{0}(q,p) + \sin^2 \tfrac{\theta}{2} W_{1}(q,p) + \frac{1}{2\pi} \cos \tfrac{\theta}{2} \sin \tfrac{\theta}{2} e^{i\phi} \widetilde{W}_{01}(q,p) \nonumber \\
                      & + \frac{1}{2\pi} \cos \tfrac{\theta}{2} \sin \tfrac{\theta}{2} e^{-i\phi} \widetilde{W}_{10}(q,p),
\end{align}
where we have defined the cross terms as follows
\begin{equation}
 \label{eq:wigner_cross}
 \widetilde{W}_{jk}(q,p) \equiv \int_{-\infty}^{\infty} dx e^{ipx}  \Psi_{j}\left(q+\tfrac{x}{2}\right)^{*}\Psi_{k}\left(q-\tfrac{x}{2}\right) .
\end{equation}
We simplify the cross terms as follows
\begin{align}
 \label{eq:wigner_crossdet}
 \widetilde{W}_{jk}(q,p) & = \int dx e^{ipx} \bigg[\sum_s \delta\left(q-\sqrt{\pi}(j+2s)+\tfrac{x}{2}\right) \bigg]\bigg[\sum_t \delta\left(q-\sqrt{\pi}(k+2t)-\tfrac{x}{2}\right) \bigg] \nonumber \\
                         & = \sum_{st} e^{i2p[q-\sqrt{\pi}(k+2t)]}\delta\left(q-\tfrac{\sqrt{\pi}}{2}(j+k+2s+2t)\right) \nonumber                                                                   \\
                         & = \sum_{st} e^{i2p[q-\sqrt{\pi}(k+2t-2s)]}\delta\left(q-\tfrac{\sqrt{\pi}}{2}(j+k+2t)\right) \nonumber                                                                   \\
                         & = \sum_{st} e^{i2p\sqrt{\pi} 2s}e^{ip\sqrt{\pi} (j-k-2t)}\delta\left(q-\tfrac{\sqrt{\pi}}{2}(j+k+2t)\right) \nonumber                                                    \\
                         & = \frac{\sqrt{\pi}}{2}\sum_{st} e^{ip\sqrt{\pi} (j-k-2t)}\delta\left(p-s\tfrac{\sqrt{\pi}}{2}\right)\delta\left(q-\tfrac{\sqrt{\pi}}{2}(j+k+2t)\right) \nonumber         \\
                         & = \frac{\sqrt{\pi}}{2}\sum_{st} (-1)^{\tfrac{s}{2}(j-k-2t)}\delta\left(p-s\tfrac{\sqrt{\pi}}{2}\right)\delta\left(q-\tfrac{\sqrt{\pi}}{2}(j+k+2t)\right) .
\end{align}

Now, combining Eq.~(\ref{eq:appwigner}) and Eq.~(\ref{eq:wigner_crossdet}), we have
\begin{align}
 W(\theta,\phi;q,p) = & \cos^2 \tfrac{\theta}{2} W_{0}(q,p) + \sin^2 \tfrac{\theta}{2} W_{1}(q,p) + \tfrac{1}{4\sqrt{\pi}} \cos\tfrac{\theta}{2} \sin\tfrac{\theta}{2}\nonumber          \\
                      & \times \bigg[e^{i\phi} \sum_{st} (-1)^{\tfrac{s}{2}(-1-2t)}\delta\left(p-s\tfrac{\sqrt{\pi}}{2}\right)\delta\left(q-\tfrac{\sqrt{\pi}}{2}(1+2t)\right) \nonumber \\
                      & + e^{-i\phi} \sum_{st} (-1)^{\tfrac{s}{2}(1-2t)}\delta\left(p-s\tfrac{\sqrt{\pi}}{2}\right)\delta\left(q-\tfrac{\sqrt{\pi}}{2}(1+2t)\right)\bigg]\nonumber       \\
 =                    & \cos^2 \tfrac{\theta}{2} W_{0}(q,p) + \sin^2 \tfrac{\theta}{2} W_{1}(q,p) \nonumber                                                                              \\
                      & + \tfrac{1}{8\sqrt{\pi}} \sin\theta \sum_{st} (-1)^{st}\left(e^{i\phi}(-1)^{\tfrac{s}{2}} +  e^{-i\phi}(-1)^{-\tfrac{s}{2}}\right) \nonumber                     \\
                      & \times \delta\left(q-\tfrac{\sqrt{\pi}}{2}(1+2t)\right) \delta\left(p-s\tfrac{\sqrt{\pi}}{2}\right).
\end{align}
Then, it follows that the Wigner function for arbitrary superpositions of GKP states is given by Eq.~(\ref{eq:wignerGKP}) in the main text.

\section*{Appendix 2}
%\addcontentsline{toc}{section}{Appendix 2}
\label{app:CO2}
The table below summarizes the estimated climate footprint of this work, including air-travel for collaboration purposes. Estimations  have  been  calculated  using  the  examples  of  ScientificCO$_2$nduct~\cite{tableCO2}.
\begin{center}
 \begin{tabular}[b]{l c}
  \hline
  \textbf{Transport}                                  &             \\
  \hline
  Total CO$_2$-Emission For Transport [$\mathrm{kg}$] & 6645        \\
  Were The Emissions Offset?                          & \textbf{No} \\
  \hline
  Total CO$_2$-Emission [$\mathrm{kg}$]               & 6645        \\
  \hline
  \hline
 \end{tabular}
\end{center}

\bibliography{./BlochGKP.bbl}

\end{document}